\documentclass[12pt]{article} 
\usepackage{graphicx,subfigure,latexsym}
\usepackage{amsmath,amssymb,amsfonts}
\usepackage{bm}
\usepackage[colorlinks=true, pdfstartview=FitV, linkcolor=red, citecolor=blue, urlcolor=blue]{hyperref}
\usepackage{cite}

 \makeatletter
    
    \@addtoreset{equation}{section}
  \makeatother
  
\newcommand{\be}{\begin{equation}}
\newcommand{\ee}{\end{equation}}
\newcommand{\bear}{\begin{eqnarray}}
\newcommand{\eear}{\end{eqnarray}}
\newcommand{\ba}{\begin{array}}
\newcommand{\ea}{\end{array}}

\def \be {\begin{equation}}
\def \ee {\end{equation}}

\def \bes {\begin{subequations}}
\def \ees {\end{subequations}}

\newcommand{\Tc}{{\text T}_c}
\newcommand{\comment}[1]{}

\def \<{\langle}
\def \>{\rangle}
\def \+{\dagger}
\def \({\left(}
\def \){\right)}
\def \[{\left[}
\def \]{\right]}

\def \vp {{\bf{p}}}

\def \vx {{\bf{x}}}

\def \vk {{\bf{k}}}
\def \vj  {{\bf{j}}}
\def \vA {{\bf{A}}}

\def \vB {{\bf{B}}}
\def \vE {{\bf{E}}}
\def \vsigma {{\boldsymbol \sigma}}

\begin{document}

\begin{titlepage}
\vfill
\begin{flushright}
{\normalsize BNL-105312-2014-JA \\ RBRC-1076 \\
RIKEN-QHP-112}\\
\end{flushright}

\vfill
\begin{center}
{\Large\bf  Chiral Magnetic Effect at Weak Coupling\\with Relaxation Dynamics }

\vskip 0.3in

\vskip 0.3in
Daisuke Satow$^{1,2}$\footnote{e-mail: {\tt daisuke.sato@riken.jp}} and 
Ho-Ung Yee$^{3,4}$\footnote{e-mail:
{\tt hyee@uic.edu}}
\vskip 0.15in

{\it $^{1}$ Department of Physics, Brookhaven National Laboratory, }\\
{\it Upton, NY-11973, USA}\\
[0.15in]

{\it $^{2}$ Theoretical Research Division, Nishina Center,
RIKEN,}\\
{\it Wako 351-0198, Japan}\\
[0.15in]

{\it $^{3}$ Department of Physics, University of Illinois, Chicago, Illinois 60607}
\\[0.15in]

{\it $^{4}$ RIKEN-BNL Research Center, Brookhaven National Laboratory,}\\
{\it Upton, New York
11973-5000}\\
[0.15in]

{\normalsize  \today
}

\end{center}

\vfill

\begin{abstract}

We provide a resolution of an old issue in weak coupling computation of the Chiral Magnetic Effect (CME) current, where a free chiral fermion theory gives two different results depending on the order of the two limits, $\omega\to 0$ (frequency) and $k\to 0$ (spatial momentum). 
We first argue based on hydrodynamics that in any reasonable interacting theory of chiral fermions the non-commutativity between the two limits should be absent, and we demonstrate this at weak coupling regime in two different frameworks: kinetic theory in the relaxation time approximation, and diagrammatic computation with resummation of damping rate. 
In the latter computation, we also show that the ``pinch'' singularity, which would make the summation of ladder diagrams necessary as in the P-even correlation function, is absent in the relevant P-odd correlation function. 
The correct value of CME current is reproduced even in the presence of relaxation dynamics in both computations.
\end{abstract}

\vfill

\end{titlepage}
\setcounter{footnote}{0}

\baselineskip 18pt \pagebreak
\renewcommand{\thepage}{\arabic{page}}
\pagebreak

\section{Introduction}

Triangle anomaly (chiral anomaly) is an interesting quantum mechanical violation of the classically conserved $U(1)$ chiral symmetry of a massless Weyl fermion in the presence of a P- and CP-odd configuration of external gauge fields coupled to the $U(1)$ current. It represents the broken conservation laws of the left and right handed currents,\footnote{This is the covariant anomaly.}
\be
\partial_\mu j^\mu_{L/R}=\mp {e^2\over 4\pi^2}\vE\cdot\vB\,,
\ee
where the sign in the right-hand side depends on the chirality of the Weyl fermion.
It gives rise to a wide range of interesting physics phenomena in different phases of QCD where one can neglect masses of light quarks and have an approximate classical 
$U(1)$ axial symmetry which suffers from triangle anomaly.
Recently, one such phenomenon originating from triangle anomaly in a quark-gluon plasma in the presence of magnetic field
has attracted much attention from both theorists and experimentalists working on heavy-ion collisions: the Chiral Magnetic Effect (CME) \cite{Kharzeev:2007tn,Kharzeev:2007jp,Fukushima:2008xe,Son:2004tq,Metlitski:2005pr} which dictates vector (axial) charge current along the direction of the magnetic field in the presence of an axial (vector) chemical potential,
\be
{\vj}_{V,A}={e^2 \over 2\pi^2} \mu_{A,V}{\vB}\,, \label{CMEintro}
\ee
in the chiral basis defined by $j_{V,A}=e(\pm j_L+j_R)$.
The interplay of the two versions of the CME (the other version interchanging axial and vector symmetries is sometimes called the Chiral Separation Effect) leads to
a gapless sound-like mode of chiral charge fluctuations, the Chiral Magnetic Wave (CMW) \cite{Kharzeev:2010gd,Newman:2005hd}. Both CME and CMW may have interesting experimental implications in off-central heavy-ion collisions \cite{Voloshin:2004vk,Bzdak:2009fc,Burnier:2011bf,Burnier:2012ae,Gorbar:2011ya,Yee:2013cya,Hongo:2013cqa,Taghavi:2013ena} where a strong and transient magnetic field is created by heavily charged ultra relativistic projectiles \cite{Kharzeev:2007jp}, some of which seem to be in line with what is observed in RHIC and LHC \cite{Abelev:2009ac,Selyuzhenkov:2011xq,Wang:2012qs,Ke:2012qb,Shou:2014cua,Ma:2014iva}.

There have been theoretical confirmations of the CME in both weak \cite{Kharzeev:2009pj} and strong coupling \cite{Yee:2009vw,Rebhan:2009vc,Gynther:2010ed,Hoyos:2011us} frameworks, and there are evidences also in lattice QCD \cite{Buividovich:2009wi,Abramczyk:2009gb,Yamamoto:2011gk,Buividovich:2013hza,Bali:2014vja}.
Also, the CME can be viewed as a leading modification of the hydrodynamic constitutive relation for the current arising from the underlying triangle anomaly \cite{Son:2009tf}. 
The modified constitutive relations up to first derivative in the local rest frame read as
\be
{\vj}_{L,R}=\sigma {\vE}-\sigma T\nabla\left(\mu\over T\right) + \sigma_\chi {\vB}+\cdots\,,
\ee
with the chiral magnetic conductivity $\sigma_\chi$ defined at zero frequency-momentum limit.
Since they are given as linear response to the small and slowly-varying magnetic field, it is natural to compute them in the appropriate Kubo formula
in a zero frequency-momentum limit \cite{Kharzeev:2009pj,Amado:2011zx}.
It is given  in terms of the P-odd part of the retarded current-current correlation function which takes a general form
\be
G_{R}^{ij,-}=i\sigma_\chi(\omega,k)\epsilon^{ijk} k_k\,,
\ee
and that the zero frequency-momentum limit of the chiral magnetic conductivity is given by the Kubo formula \cite{Kharzeev:2009pj,Amado:2011zx}
\be
\sigma_\chi=\lim_{k\to 0}\lim_{\omega\to 0} {-i\over 2 k_k}\sum_{ij}\epsilon^{ijk}G^{ij,-}_R\,.\label{weaksigmachi}
\ee

A puzzling problem appeared in a weak coupling computation of (\ref{weaksigmachi}) in Ref.\cite{Kharzeev:2009pj} where interchanging the order of the two limits, $\omega\to 0$ and $k\to 0$, gives
different results,
\be
\lim_{k\to 0}\lim_{\omega\to 0} {-i\over 2 k_k}\sum_{ij}\epsilon^{ijk}G^{ij,-}_R\neq\lim_{\omega\to 0}\lim_{k\to 0} {-i\over 2 k_k}\sum_{ij}\epsilon^{ijk}G^{ij,-}_R\,,\label{notequal}
\ee
where the latter gives only 1/3 of the first expression which is the right value $\sigma_\chi={e^2\mu/4\pi^2}$.
The computation in Ref.~\cite{Kharzeev:2009pj} was done in a free fermion theory, and the same non-commutativity of the two limits was also found in the chiral kinetic theory \cite{Son:2012wh,Stephanov:2012ki} with free fermions \cite{Son:2012zy}.
On the other hand, a computation in the strong coupling regime in the AdS/CFT correspondence was performed in Ref.~\cite{Yee:2009vw} using the latter limit, which found the correct
expected result and it can be shown that there exists no such non-commutativity in the two limits in the strong coupling regime \footnote{We thank Shu Lin for pointing this out to us.}.

This has led to several speculations on why and how this non-commutativity in the weak coupling (free theory) computation arises.
Refs.~\cite{Kharzeev:2009pj,Hou:2011ze} suggest that it might be an artifact of free theory and will disappear when the interaction effect is taken into account.
Another speculation is that this difference between the two limits might be related to a difference between covariant and consistent anomaly \cite{Wess:1971yu}, since the latter has in fact 1/3 smaller coefficient from the former. 
Also, based on the observation that the difference is absent in the strong coupling limit, one may have another speculation that the difference might depend on the coupling constant and interpolates from 2/3 to 0 as one varies the coupling strength from zero to infinity.
Up to our knowledge, there has not been a clear resolution of the issue discussed in literature, and these speculations have remained neither verified nor disproved.

In this work, we provide a resolution of this puzzle by showing that the first speculation is correct:
the non-commutativity of the two limits in fact is an artifact of the free fermion theory, and is absent in any interacting theory that has a smooth infrared regime at sufficiently small frequency-momentum. 
We first give an argument for this in hydrodynamics, and demonstrate it in two different frameworks: chiral kinetic theory \cite{Son:2012wh,Stephanov:2012ki} with relaxation dynamics, and diagrammatic computation with resummation of damping rate. 
We also confirm that the right magnitude of the CME in these frameworks is obtained in the presence of damping/relaxation dynamics.

This paper is organized as follows.
In the next section, we present our argument based on a general consideration of hydrodynamics which should be able to capture the correct infrared behavior of any correlation functions at sufficiently small frequency-momentum regime:
if there existed such a non-commutativity between $\omega\to 0$ and $k\to 0$ limits, the hydrodynamics should be able to see it, whereas we show that there is no such behavior in hydrodynamic regime. 
To corroborate this in explicit microscopic computations, we compute the necessary P-odd correlation functions in two microscopic frameworks of weakly interacting chiral fermion theory:
in section~\ref{sec:kinetic}, we compute
in the chiral kinetic theory \cite{Son:2012wh,Stephanov:2012ki} now with a non-zero collision term in the relaxation time approximation, and in section~\ref{sec:diagram}, we perform the diagrammatic computation with resummation of damping rate~\cite{Lebedev:1989ev,Jeon:1994if,ValleBasagoiti:2002ir, Gagnon:2006hi,Hidaka:2011rz, Satow:2013oya, Blaizot:2014hka} in the propagators. 
In the latter case, we also show that the usual ``pinch'' singularity~\cite{Lebedev:1989ev,Jeon:1994if,ValleBasagoiti:2002ir, Gagnon:2006hi, Hidaka:2011rz, Satow:2013oya, Blaizot:2014hka} that necessitates an infinite ladder resummation of the vertex at zero momentum limit in fact disappears for the P-odd correlator. 
In both microscopic frameworks, we indeed observe that the non-commutativity of the two limits disappears, and the two expressions in (\ref{notequal}) give the same expected result. 
We give a summary and concluding remarks
 in section~\ref{sec:summary}.

\section{Hydrodynamics}
\label{sec:hydro}

In this section, we present our first argument, based on hydrodynamics, on the absence of non-commutativity between $\omega\to 0$ and $k \to 0$ limits in the Kubo formula of CME, in any reasonably interacting theory which reduces to hydrodynamics at sufficiently low energy.

To highlight salient features of our argument, let us first consider a well-known example where 
there is indeed such a non-commutativity between $\omega\to 0$ and $k\to 0$ limits: longitudinal current-current retarded correlators.
Recall that one can extract the retarded current-current correlators from the linear response of the current $j^\mu$ to an external gauge potential $A_\nu$,
\be
j^\mu(k)=-G_R^{\mu\nu}(k) A_\nu(k)\,,\label{defretarded}
\ee
where $k_\mu=(\omega,\vk)$ denotes a frequency-momentum four vector.
In the hydrodynamic regime, one can obtain the response current $j^\mu$ by solving hydrodynamic equations of motion under the influence of electric and magnetic fields arising from the gauge potential.
The constitutive relation for the current gives
\be
{\vj}=-D \nabla j^0+\sigma {\vE}+\cdots\,,
\ee
with a diffusion constant $D$ and a conductivity $\sigma$, where we neglected higher derivative corrections.
With the conservation of the current    
\be 
\partial_0 j^0 +\nabla\cdot {\vj}=0\,,
\ee
and neglecting couplings to the energy-momentum sector which is valid for linearized fluctuations out of neutral plasma, one has a self-closed dynamical system.
For our purpose, let us choose ${\vk}=k \hat z$ and look at longitudinal components of the currents only, $j^0$ and $j^3$, for which the above equations become
\be
\label{eq:current}
j^3=-ikD j^0 +\sigma \left(i\omega A_3+ik A_0\right)\,,\quad
-i\omega j^0+i k j^3=0\,,
\ee
where we used $\vE=-\partial_0 \vA+\nabla A_0$. Solving them gives
\be
\label{eq:j-longitudinal}
j^0={i\sigma\left(k^2 A_0+k\omega A_3\right)\over \omega+iD k^2}\,,
\quad j^3={i\sigma\left(\omega k A_0+\omega^2 A_3\right)\over \omega+iD k^2}\,,
\ee
and comparing with the definition of the retarded correlators (\ref{defretarded}), we get
\be
G_R^{00}={-i\sigma k^2\over\omega+iDk^2}\,,\quad
G_R^{03}=G_R^{30}={-i\sigma \omega k\over\omega+iDk^2}\,,\quad
G_R^{33}={-i\sigma \omega^2\over\omega+iDk^2}\,,\label{Gmunu}
\ee
which satisfy the Ward identity $-i\omega G_R^{0\mu}+ik G_R^{3\mu}=0$ ($\mu=0,3$).
One can then think of a Kubo formula for the conductivity
\be
\label{eq:Kubo-sigma}
\sigma=\lim_{\omega\to 0}\lim_{k\to 0} {i\over \omega}G_R^{33}(\omega,k)\,,
\ee
where it is clear that the other ordering of the two limits would give a different result,
\be
\lim_{k\to 0}\lim_{\omega\to 0} {i\over \omega}G_R^{33}(\omega,k)=0\,.
\ee
The two lessons we learn from this example are

1) Hydrodynamics is able to identify a non-commutativity between $\omega\to 0$ and $k\to 0$ limits in the retarded correlation functions, if it exists. This is because hydrodynamics is a universal effective theory valid for sufficiently small $\omega$ and $k$, so that the two limits $\omega\to 0$ and $k \to 0$ 
that one is considering are both describable within the regime of hydrodynamics. If there is such a non-commutativity, it should be detectable within the framework of hydrodynamics. 

2) The non-commutativity arises due to the presence of a hydrodynamic, ``massless'' pole at $\omega=-iDk^2$ which gives rise to a particular long wavelength IR dynamics (a charge diffusion in this example).
This IR mode gives rise to the above non-analytic IR singularity in $G_R$ near the point $(\omega,k)=(0,0)$.
Without such an IR mode, the point $(\omega,k)=(0,0)$ would not be singular and the non-commutativity would be absent.

Guided by these lessons, let us then consider the P-odd part of the retarded correlation functions responsible for CME in the hydrodynamic regime. For simplicity we consider the case of single right-handed Weyl fermion species. The hydrodynamic constitutive relation including CME reads as
\be
\vj=-D\nabla j^0+\sigma\vE+{e^2\mu\over 4\pi^2}\vB+\cdots\,,
\label{eq:anomaly}
\ee
where $\mu$ is the background chemical potential, and $\vB=\nabla\times\vA$ is the magnetic field. 
As before, we choose $\vk=k\hat z$.
Then, we obtain the equation governing the longitudinal part of the current, and it is identical to the result without CME, the first equation in Eq.~(\ref{eq:current}). 
In conjunction with the anomaly relation
\be
\partial_0 j^0+\nabla\cdot\vj={e^3\over 4\pi^2} \vE\cdot\vB\,,
\ee
which gives us
\be
-i\omega j^0+ikj^3=0\,,
\ee
up to linear order in gauge potential, the longitudinal retarded correlation functions are identical
to those in the previous example without CME.
The CME affects the transverse part of the correlation functions as
\be
j^1=i\omega \sigma A^1 -i{e^2\mu\over 4\pi^2}k A^2\,,\quad 
j^2=i\omega \sigma A^2 +i{e^2\mu\over 4\pi^2}k A^1\,, 
\ee
where the second terms in the right-hand sides are the P-odd contribution from CME. 
This gives us the transverse part of the retarded
correlation functions as
\be
G_{R}^{T,ij}=-i\omega\sigma \left(\delta^{ij}-{k_i k_j\over k^2}\right)+i{e^2\mu\over 4\pi^2}\epsilon^{ijk} k_k\,.\label{CMEkubo}
\ee
From this equation, one obtains the Kubo formula for the chiral magnetic conductivity~\cite{Kharzeev:2009pj,Amado:2011zx}, Eq.~(\ref{weaksigmachi}), and it is clear that the other ordering of the two limits gives us the same result,
\be
\sigma_\chi=\lim_{k\to 0}\lim_{\omega\to 0} {-i\over 2 k_k}\sum_{ij}\epsilon^{ijk}G^{T,ij}_R
=\lim_{\omega\to 0}\lim_{k\to 0} {-i\over 2 k_k}\sum_{ij}\epsilon^{ijk}G^{T,ij}_R={e^2\mu\over 4\pi^2}\,.
\ee
where there is no sum over the index $k$ in the last expression.
This is logically related to the fact that there is no hydrodynamic pole in the P-odd, transverse part of the retarded current-current correlation functions, as can be seen from Eq.~(\ref{CMEkubo}).

In light of our discussion above, we can understand why the previous computations in the free fermion theory possesses the non-commutativity of $\omega\to 0$ and $k\to 0$ limits in the Kubo formula for CME: the free fermion theory simply does not have a hydrodynamic regime. Once we introduce interactions, however weak it is, the theory is expected to have a hydrodynamic regime at sufficiently long wavelength regime, and the non-commutativity in the Kubo formula for CME is expected to disappear.
The origin of the non-commutativity in the free theory can be traced to the existence of IR singularity due to the absence of interaction. 
It is natural to expect this IR singularity to be removed by interactions.

In the following sections, we will confirm this expectation explicitly in two microscopic frameworks: kinetic theory and diagrammatic computation with relaxation/damping rate included.

\section{Kinetic Theory} 
\label{sec:kinetic}
Let us start with a brief summary of recently developed kinetic theory of chiral fermions~\cite{Son:2012wh,Stephanov:2012ki}. 
Since we are interested in a system of finite temperature and chemical potential, we
need to consider both particles and anti-particles 
in the description, which is a modest extension of the previous discussions in literature \cite{Manuel:2013zaa}.
A computation of retarded current-current correlation functions in kinetic theory of free chiral fermions has been done before in Ref.~\cite{Son:2012zy,Manuel:2013zaa}, and our computation is a simple generalization of it, now including relaxation dynamics in the collision term in the Boltzmann equation. 

We consider a single species of right-handed Weyl fermion field whose quantization gives us right-handed fermionic particles with helicity $h=+1/2$ and left-handed anti-particles with $h=-1/2$.
They carry $U(1)$ charge of $+1$ and $-1$ respectively which couples to an external gauge potential $A_\mu$. 
Introducing distribution functions $f_\pm({\bf x},{\bf  p},t)$ for particles and anti-particles with momentum $\vp$ at space-time point $(\vx,t)$, the Boltzmann equations for them takes the form
\be
{\partial f_\pm\over\partial t}+\dot{\bf x}_\pm \cdot {\partial f_\pm\over\partial {\bf x}}+\dot{\bf p}_\pm \cdot {\partial f_\pm\over\partial {\bf p}}={\cal C}_\pm [f_+,f_-]\,,\label{boltzmann}
\ee 
where $(\dot{\bf x}_\pm,\dot{\bf p}_\pm)$ are obtained from the equations of motion of a single particle and anti-particle that have to be specified by the underlying theory, and ${\cal C}_\pm$ are the collision terms reflecting interactions of the particles/anti-particles.

The single particle/anti-particle equations of motion are derived from the action
\be 
S_\pm=\int dt\left[ {\bf p}\cdot\dot{\bf x}\pm e{\bf A}({\bf x},t)\cdot \dot{\bf x}\pm eA_0({\bf x},t)-{\cal E}({\bf p},{\bf x},t) \mp {\bf \cal A}_p({\bf p})\cdot\dot{\bf p} \right]\,,\label{action}
\ee
where ${\cal E}$ is the energy of the particle, and ${\bf \cal A}_p$ is the Berry phase in momentum space~\cite{Stephanov:2012ki, Son:2012wh}.
We do not need an explicit form of ${\bf \cal A}_p$, except its curvature introduced later.
We note that this term is from the Berry connection in the momentum space that arises from
the chirality (helicity) projection of particle/anti-particle wave functions. 
It has been shown that this term can reproduce
all the salient features of chiral anomaly within the present
framework of kinetic theory. 
Note that particles and anti-particles have the opposite signs of Berry connection, reflecting their opposite chiralities. Note also that we use the same energy ${\cal E}$ for particles and anti-particles, which will be explained more clearly below.
The equations of motion read from Eq.~(\ref{action}) as
\bear
\sqrt{G}\dot{\bf x}_\pm&=&{\partial {\cal E}\over\partial{\bf p}}\mp{\partial{\cal E}\over\partial{\bf x}}\times{\bf b}
+e{\bf E}\times{\bf b}+e{\bf B}\left({\partial {\cal E}\over \partial{\bf p}}\cdot{\bf b}\right)\,,\nonumber\\
\sqrt{G}\dot{\bf p}_\pm&=&-{\partial {\cal E}\over\partial{\bf x}}\pm e{\bf E}\pm e{\partial{\cal E}\over\partial{\bf p}}\times{\bf B}
+e{\bf b}\left(-{\partial {\cal E}\over \partial{\bf x}}\cdot{\bf B}\pm e{\bf E}\cdot{\bf B}\right)\,,
\eear
where 
\be
{\bf b}={\bf \nabla}_p\times{\bf\cal A}_p={\hat{\bf p}\over 2 |{\bf p}|^2}\,
\ee
is the Berry curvature of a monopole shape, and 
\be
\sqrt{G}=1+e{\bf b}\cdot{\bf B}\,
\ee
is the modified phase space measure due to the Berry curvature. For the energy $\cal E$, we use the form
suggested in Ref.~\cite{Son:2012zy} based on the Lorentz invariance of the kinetic theory~\cite{Chen:2014cla,Manuel:2014dza},
\be
{\cal E}=|{\bf p}|-e{{\bf B}\cdot\hat{\bf p}\over 2|{\bf p}|}\,,\label{energy}
\ee
where the second term may be interpreted as an interaction between spin-induced magnetic moment and the magnetic field. The sign of this interaction is determined by the product of helicity $h$ and the $U(1)$ charge, and it is the same for both particles and anti-particles, explaining our use of the same $\cal E$ for them. From the distribution functions, we compute the current density as follows~\cite{Son:2012zy}:
\bear
j^0({\bf x},t)&=&e\int{d^3{\bf p}\over (2\pi)^3}\sqrt{G}\left(f_+({\bf p},{\bf x},t)-f_-({\bf p},{\bf x},t)\right)\,,\nonumber\\
{\bf j}({\bf x},t)&=&-e\int {d^3{\bf p}\over (2\pi)^3}\Bigg({\cal E}{\partial f_+\over\partial{\bf p}}+e{\cal E}\left({\bf b}\cdot{\partial f_+\over\partial{\bf p}}\right) {\bf B}+{\cal E} {\bf b}\times {\partial f_+\over\partial{\bf x}}\nonumber\\
&&-{\cal E}{\partial f_-\over\partial{\bf p}}-e{\cal E}\left({\bf b}\cdot{\partial f_-\over\partial{\bf p}}\right) {\bf B}+{\cal E} {\bf b}\times {\partial f_-\over\partial{\bf x}}
-e{\bf E}\times{\bf b}\left(f_++f_-\right)\Bigg)\,,
\label{current}
\eear
It is important to have the correct phase space measure $\sqrt{G}$ in the above expressions:
only with this definition of the current, the correct anomaly relation (\ref{eq:anomaly}) is reproduced. 

What remains to be specified is the collision terms, and we will use a relaxation time approximation.
An important requirement on the collision terms is that they should respect the conserved quantities of the system such as energy-momentum and global charges. Although we neglect the interplay with energy-momentum fluctuations for simplicity, the local conservation of our $U(1)$ global charge should be 
imposed strictly on the collision terms for a consistent treatment of charge fluctuations we are going to study.
This means that
\be
\int {d^3{\bf p}\over(2\pi)^3} \,\sqrt{G}\left( {\cal C}_+[f_+,f_-]-{\cal C}_-[f_+,f_-]\right)=0\,,\label{constraint}
\ee
has to be satisfied for any $f_\pm$.
As we will be interested in a linearized fluctuation of distribution functions from their equilibrium values
\be
n_\pm(E)\equiv {1\over e^{\beta(E\mp\mu)}+1}\,,
\ee
it is convenient to parameterize the fluctuation as
\be
f_\pm=n_\pm +\beta n_\pm (1-n_\pm)\left({e{\bf B}\cdot\hat{\bf p}\over 2 |{\bf p}|}+h_\pm\right)\,,
\ee
where we define the fluctuations $h_\pm$ after pulling out explicitly the effect of the energy shift due to the spin-magnetic interaction in (\ref{energy}) following Ref.~\cite{Son:2012zy}. The relaxation dynamics dampens the fluctuations $h_\pm$ up to
the constraint (\ref{constraint}), and it is important to separate as above  the effect of the spin-magnetic energy shift, which is not subject to the relaxation dynamics, in order to get the correct magnitude of CME.  
Then the collision terms in the relaxation time approximation are
\be
{\cal C}_\pm=-{1\over\tau}\beta n_\pm (1-n_\pm)\left(h_\pm\mp \delta \mu\right)\,,
\ee
where $\tau$ is the relaxation time of the fermion. 
If we consider gauge theory, $\tau^{-1}$ is of order $e^4T \ln (1/e)$ when $\mu\lesssim T$.
$\delta\mu$, interpreted as a local shift of chemical potential in the equilibrium with a given charge fluctuation, is given by solving Eq.~(\ref{constraint}) as
\be
\delta\mu={1\over \chi }\int {d^3{\bf p}\over (2\pi)^3}\,\sqrt{G}\left(\beta n_+(1-n_+) h_+-\beta n_-(1-n_-) h_-\right)\,,\label{deltamu}
\ee
where \be\chi\equiv{T^2\over 6}+{\mu^2\over 2\pi^2}\,,\ee
and we neglected contributions which are quadratic or higher in ${\bf B}$ as we work only linearly in perturbations. 
Since $\delta\mu$ involves an integral of $h_\pm$ in the momentum space, the Boltzmann equation with the above collision terms is an integro-differential equation for the fluctuations $h_\pm$.

To compute retarded current-current correlation functions within the framework of kinetic theory, we first obtain 
the linearized response of the distribution functions to the external gauge field strengths ${\bf E}$ and ${\bf B}$ by solving the Boltzmann equations (\ref{boltzmann}).
Linearizing (\ref{boltzmann}) in $h_\pm$ and $({\bf E},{\bf B})$ gives after some algebra,
\be
{\partial h_\pm\over\partial t}+\hat{\bf p}\cdot {\partial h_\pm\over\partial {\bf x}}= -{e\over 2|{\bf p}|}{\partial\over\partial t}({\bf B}\cdot\hat{\bf p})\pm e{\bf E}\cdot\hat{\bf p} -{1\over\tau}\left(h_\pm\mp\delta\mu\right)\,,
\ee
where $\delta\mu$ is given in (\ref{deltamu}) and $\hat\vp\equiv \vp/|\vp|$. 
It is straightforward to solve for $h_\pm$ assuming a definite frequency-momentum $(\omega,{\bf k})$ to find
\be
h_\pm={e{i\omega}{\bf B}\cdot\hat{\bf p}/(2|{\bf p}|)\pm e{\bf E}\cdot\hat{\bf p}\pm{\tau^{-1}}\delta\mu
\over -i\omega+i\hat{\bf p}\cdot {\bf k}+{\tau^{-1}}}\,,\label{sol1}
\ee
where $\delta\mu$, and hence $h_\pm$, should be found by solving the consistency equation obtained
by inserting (\ref{sol1}) into (\ref{deltamu}). 
Here we note that the condition $k \ll p$ is implicitly imposed so that the kinetic theory is valid, since the separation of the scale of the particle ($p$) and the field ($k$) is necessary for the description of the kinetic theory.
Since $p \sim T,\mu$, which can be confirmed by looking at the $|\vp|$ integrations in the derivation of Eq.~(\ref{deltamu}), the condition above is reduced to $k \ll T,\mu$. 
Performing necessary radial and angular integrations of ${\bf p}$, we arrive at a result
\bear
\delta\mu&=&{e\over 1-G(\omega,k)/(2ik\tau)}\int{d\Omega_{\hat{\bf p}}\over (4\pi)}\,{{\bf E}\cdot\hat{\bf p}\over 
  -i\omega+i\hat{\bf p}\cdot {\bf k}+{\tau^{-1}}}\label{deltamuresult}\\
  &=&{e\over 1-G(\omega,k)/(2ik\tau)}{{\bf E}\cdot{\bf k}\over ik^2}\left(1+{\left(\omega+{i\tau^{-1}}\right)\over2 k}G(\omega,k)\right)
,\nonumber
\eear
where $k\equiv |{\bf k}|$ and,
\bear
G(\omega,k)&=&\log\left({\omega-k+{i\tau^{-1}}\over
\omega+k+{i\tau^{-1}}}\right).\nonumber
\eear
In deriving this, we used the fact
\be
\int{d\Omega_{\hat{\bf p}}\over (4\pi)}\,{{\bf B}\cdot\hat{\bf p}\over 
  -i\omega+i\hat{\bf p}\cdot {\bf k}+{\tau^{-1}}}=0\,,
  \ee
which can be shown from ${\bf B}=i{\bf k}\times{\bf A}$ in the frequency-momentum space.
Equations (\ref{sol1}) and (\ref{deltamuresult}) constitute the solution for linearized response of the distribution functions to the external gauge fields perturbations.

Having obtained the distribution functions, we need to do straightforward but tedious computation to obtain the response current by inserting the distribution functions into the current expression (\ref{current}).
The induced charge density in linear order in perturbation is
\be
\delta j^0
=e\chi\delta\mu\,,
\ee
where we see that $e\chi$ is interpreted as the charge susceptibility.
The current is obtained from Eq.~(\ref{current}) after a sizable computation,
\begin{align}
\begin{split}
\delta{\bf j}&=e^2{\mu\over 4\pi^2} \left[{\bf B}
+\int{d\Omega_{\hat{\bf p}}\over (4\pi)}\,{i\omega ({\bf B}\cdot\hat{\bf p})\hat{\bf p}
\over -i\omega+i\hat{\bf p}\cdot{\bf k}+{\tau^{-1}}}
-{i}\int{d\Omega_{\hat{\bf p}}\over (4\pi)}
{(\hat{\bf p}\times{\bf k})({\bf E}\cdot\hat{\bf p})\over-i\omega+i\hat{\bf p}\cdot{\bf k}+{\tau^{-1}}}\right]
\\
&~~~+e^2\int{d\Omega_{\hat{\bf p}}\over (4\pi)}\, 
{\chi ({\bf E}\cdot\hat{\bf p})\hat{\bf p}
\over -i\omega+i\hat{\bf p}\cdot{\bf k}+{\tau^{-1}}}
+ {e\over\tau}\chi{{\bf k}\over  i k^2}\left(1+{\left(\omega+{i\tau^{-1}}\right)\over 2k}G(\omega,k)\right) \delta\mu\,,
\end{split}
\end{align}
where the first line is P-odd response of the current responsible for CME. 
We dropped the terms of higher order in derivatives due to $k\ll T, \mu$ which is beyond the subject of this paper.
We note that the dominant contribution to such higher order terms comes from the soft region $p\sim eT$ where the screening effect is not negligible, so we would need the Hard Thermal Loop resummation~\cite{Pisarski:1988vd, Braaten:1989kk, Braaten:1989mz, Braaten:1990it} if we would want to study it.
The conservation law $-i\omega\delta j^0+i{\bf k}\cdot\delta{\bf j}=0$, which should hold in linear order in the gauge fields, is indeed satisfied by the above results in somewhat nontrivial way, which would not be true without the $\delta\mu$ term in the collision terms. 

From ${\bf E}=i\omega {\bf A}+i{\bf k} A_0$, ${\bf B}=i{\bf k}\times{\bf A}$ and the above expressions of the current in terms of ${\bf E}$ and ${\bf B}$, we can easily read off the retarded
current-current correlation functions from kinetic theory.
After performing necessary angular integrations, we have
\bear
G_R^{00}&=&-{e^2\chi\over 1-G(\omega,k)/(2ik\tau)}
\left(1+{\left(\omega+{i\tau^{-1}}\right)\over2 k}G(\omega,k)\right)\,,\nonumber\\
G_R^{0i}=G_R^{i0}&=&-{e^2\chi\over 1-G(\omega,k)/(2ik\tau)}
{\omega k^i\over k^2}\left(1+{\left(\omega+{i\tau^{-1}}\right)\over2 k}G(\omega,k)\right)\,,\nonumber\\
G_R^{ij,+}&=& e^2\chi\int {d\Omega_{\hat{\bf p}}\over (4\pi)}{-i\omega\,\hat p^i \hat{p}^j\over -i\omega+i\hat{\bf p}\cdot{\bf k}+{\tau^{-1}}} \nonumber
\\&+& {ie^2\omega\over\tau}\chi{1\over 1-G(\omega,k)/(2ik\tau)}
{k^ik^j\over k^4}
\left(1+{\left(\omega+{i\tau^{-1}}\right)\over2 k}G(\omega,k)\right)^2
\label{eq:G-even},
 \eear
which are the P-even parts of the correlation functions.
Though they are not related to the CME, which is the subject in this paper, we present the above expressions for completeness and also to check whether the Einstein relation is satisfied.
We see that Eq.~(\ref{eq:G-even}) reproduces the result of the hard thermal/dense loop approximation~\cite{Weldon:1982aq, Frenkel:1989br, Braaten:1990az, Altherr:1992mf, Vija:1994is} when we take $1/\tau\to 0$.
By taking small $\omega, k\ll 1/\tau$ limits of $G^{00}_R$,
\be
G^{00}_R\approx {-ie^2\chi {\tau} k^2/3\over \omega+i{\tau}k^2/3}\,,
\ee and comparing with the hydrodynamics expression Eq.~(\ref{Gmunu}), we confirm the matching to the hydrodynamic regime with
\be
\label{eq:sigma-kinetic}
D={\tau\over 3}\,,\quad \sigma=e^2\chi {\tau\over 3}\,.
\ee
We see that they satisfy the Einstein relation $\sigma=e^2\chi D$.

The P-odd part of our interest is given by
\bear
G_R^{ij,-}&=&
ie^2{\mu\over 4\pi^2}\epsilon^{ijk}k^k
-e^2{\mu\omega\over 4 \pi^2}\int{d\Omega_{\hat{\bf p}}\over (4\pi)}
{\epsilon^{ilk}\hat{p}^j - \epsilon^{jlk}\hat{p}^i\over
-i\omega +i\hat{\bf p}\cdot{\bf k}+{\tau^{-1}}}\hat{p}^l{k}^k\nonumber\\
&=&ie^2{\mu\over 4\pi^2}\epsilon^{ijk}{ k}^k
\left(1-F(\omega,k)\right)\,,\label{p-odd}
\eear
where 
\be
F(\omega,k)={\omega\over k}\left({\left(\omega+{i\tau^{-1}}\right)\over k}-\left({k^2-\left(\omega+{i\tau^{-1}}\right)^2\over 2 k^2}\right)G(\omega,k)\right)\,.\label{F}
\ee
Equation (\ref{p-odd}) reproduces the result in Ref.~\cite{Son:2012zy} in the collision-less limit $1/\tau\to 0$, and our result shows a small improvement from that, replacing the denominator in the angular integration in Eq.~(\ref{p-odd}) from $-i\omega +i\hat{\bf p}\cdot{\bf k}$ with $-i\omega +i\hat{\bf p}\cdot{\bf k}+{1/\tau}$.
It is this small modification by the relaxation time that changes the behavior of $G_R^{ij,-}$ at small $\omega, k$ completely, from the collision-less, IR-singular one to a smooth, hydrodynamic one
without any IR singularity, and the non-commutativity between the two limits, $k\to 0$ and $\omega\to 0$, is absent for $F(\omega,k)$, and $F(\omega,k)\to 0$ without any ambiguity.
This change can be already seen in the integral expression in (\ref{p-odd}) : by having a relaxation time
in the denominator, the integrand does not have a pole near $\omega,k\to 0$ limits, and the result
becomes smooth around that point. This is an explicit demonstration that in a theory with finite interactions,
the Kubo formula for CME does not care the order of the limits between $\omega\to 0$ and $k\to 0$.
\begin{figure}[t]
	\centering
	\includegraphics[width=14cm]{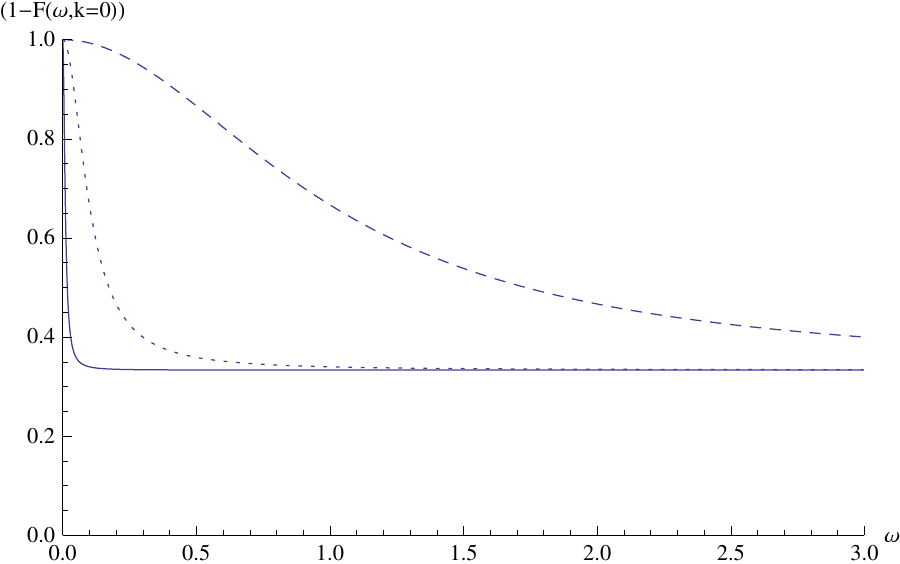}
		\caption{The plot of $(1-F(\omega,k=0))$ as a function of $\omega$ near $\omega=0$ for different values of $\tau$: $\tau^{-1}=1$ (dashed), $\tau^{-1}=0.1$ (dotted), and $\tau^{-1}=0.01$ (solid).
		The unit of $\tau^{-1}$ and $\omega$ is arbitrary.
		\label{fig1}}
\end{figure}

To visualize the effect of relaxation time in smoothing the infrared singularity, we plot $(1-F(\omega,k=0))$ as a function of $\omega$ for 
different relaxation time $\tau$ in Figure \ref{fig1}.
We see that the function rises from the value 1/3 to 1 in an interval near $\omega=0$ whose width increases as $1/\tau$ increases.
Therefore, the discontinuous jump at $\omega=0$ in the singular free theory limit of $1/\tau\to 0$
is smoothened by the relaxation dynamics, and the non-commutativity between $\omega\to 0$ and $k\to 0$ limits disappears.

\section{Diagrammatic Computation}
\label{sec:diagram}
\begin{figure}[t]
	\centering
	\includegraphics[width=8cm]{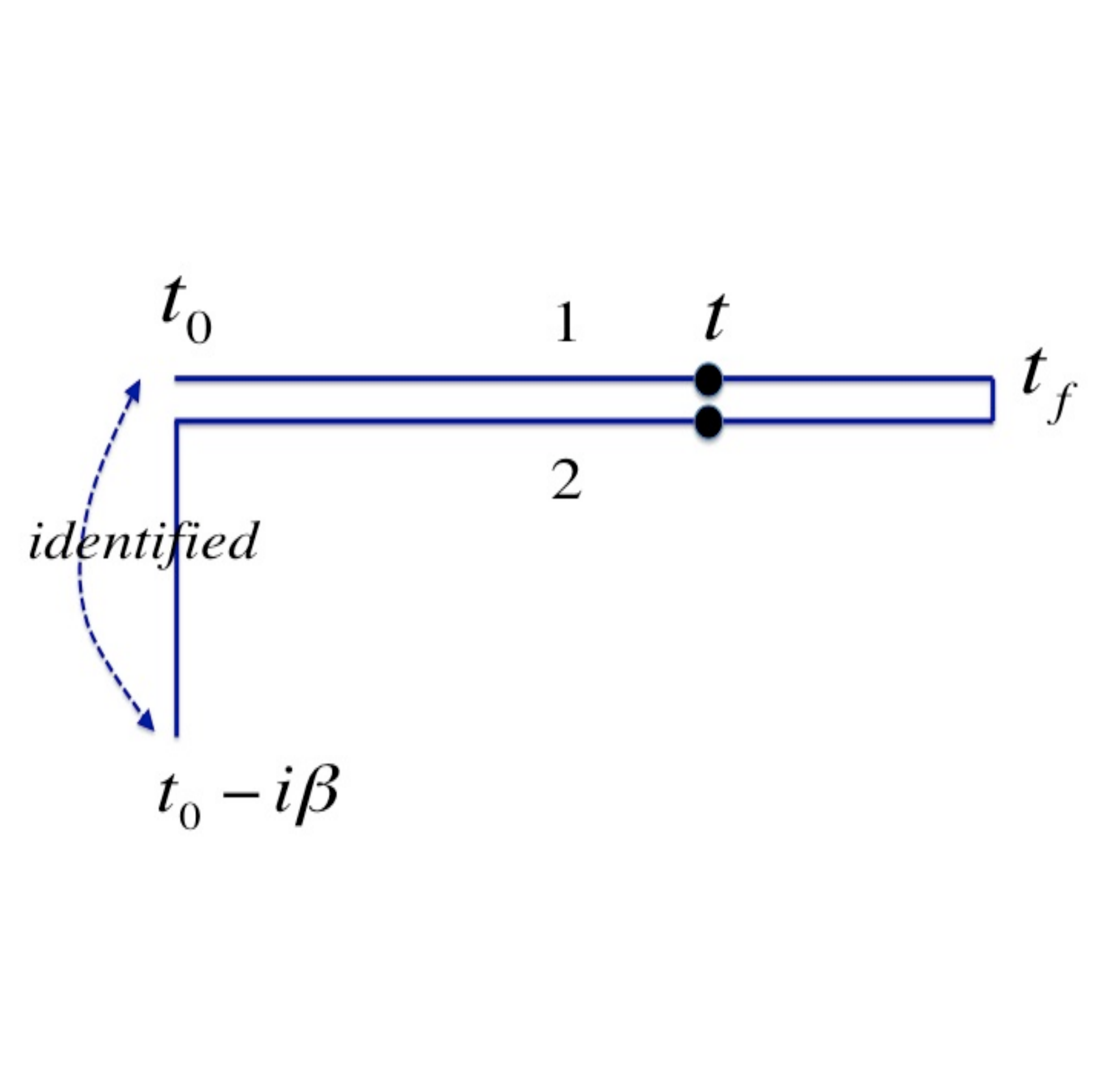}
		\caption{The Schwinger-Keldysh contour appropriate for computing real-time retarded response functions at finite temperature.\label{fig:contour}}
\end{figure}

Since we need to compute the retarded correlation functions at finite temperature and density diagrammatically, it is convenient to work in the Schwinger-Keldysh contour~\cite{Schwinger:1960qe,Keldysh:1964ud} as shown in Figure \ref{fig:contour}.
The free chiral (Weyl) fermion action is
\be
{\cal L}=i\psi^\dagger \sigma^\mu\left(\partial_\mu-i eA_\mu\right)\psi\,,
\ee
where $\sigma^\mu=({\bf 1},\vsigma)$ and $A_\mu$ is the external gauge potential.
The imaginary time contour on the far left provides the thermal ensemble, and the total action on lines 1 and 2 is
\bear
{\cal L}_{SK}&=&{\cal L}_1-{\cal L}_2=i\psi_1^\dagger \sigma^\mu\left(\partial_\mu-i eA_\mu\right)\psi_1-i\psi_2^\dagger \sigma^\mu\left(\partial_\mu-i eA_\mu\right)\psi_2\nonumber\\
&=& i\left(\psi^\dagger_a\sigma^\mu\partial_\mu\psi_r+\psi^\dagger_r\sigma^\mu\partial_\mu\psi_a
\right) +e\left(\psi_a^\dagger\sigma^\mu\psi_r+\psi_r^\dagger\sigma^\mu\psi_a\right)A_\mu\,,\label{SKaction}
\eear
where $\psi_{1,2}$ are the fields on the lines 1 and 2 respectively, and we have introduced ``ra''-basis,
\be
\psi_r\equiv{1\over 2}\left(\psi_1+\psi_2\right)\,,\quad \psi_a\equiv \psi_1-\psi_2\,.
\ee
In terms of (1,2) or (r,a) fields, any type of real-time correlation functions can be expressed.
The fermion two-point functions (propagators) in the (r,a)-basis are defined by
\bear
S_{ra}(x,y)&=&\langle \Tc \psi_r(x)\psi^\dagger_a(y)\rangle_{\rm SK}=\theta(x^0-y^0)\langle \{\psi(x),\psi^\dagger(y)\}\rangle\,,\nonumber\\
S_{ar}(x,y)&=&\langle \Tc \psi_a(x)\psi^\dagger_r(y)\rangle_{\rm SK}=-\theta(y^0-x^0)\langle \{\psi(x),\psi^\dagger(y)\}\rangle\,,\nonumber\\
S_{rr}(x,y)&=&\langle \Tc \psi_r(x)\psi^\dagger_r(y)\rangle_{\rm SK}={1\over 2}\langle [\psi(x),\psi^\dagger(y)]\rangle\,.
\eear
Here $\Tc$ is a contour-ordering operator.
Note that $S_{aa}$ vanishes identically, and $S_{ra} (S_{ar})$ is $i$ times of the usual retarded (advanced) correlator.

\begin{figure}[t]
	\centering
	\includegraphics[width=8cm]{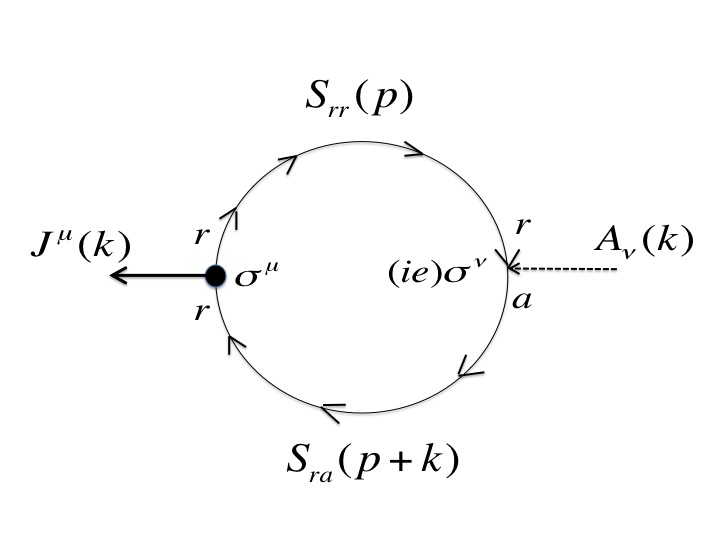}\includegraphics[width=8cm]{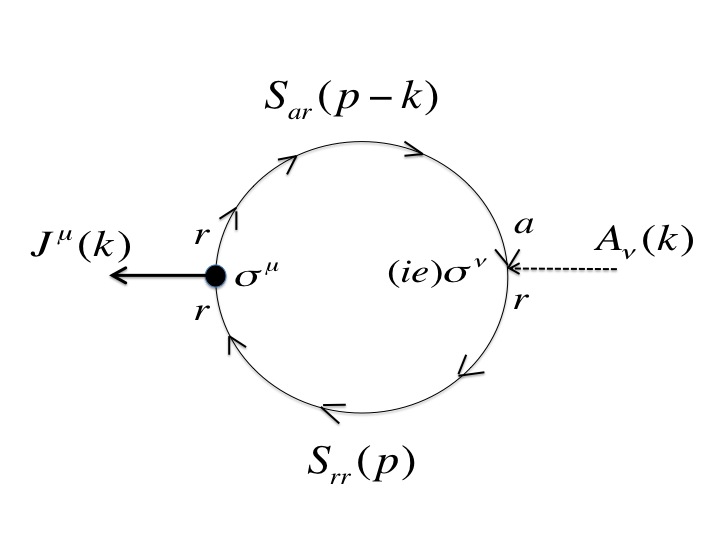}
		\caption{The diagrams responsible for the retarded response of the current $J^\mu$ to one external gauge potential.\label{fig3}}
\end{figure}

We are interested in computing the expectation value of current $j^\mu=e\psi_r^\dagger\sigma^\mu\psi_r$ in linear response to the external gauge potential $A_\mu$, and it is easy to do Feynman diagram computation for it, based on our action (\ref{SKaction}). 
There are two terms depicted in Figure \ref{fig3}, which read
\be
j^\mu(k)=(-1)ie^2A_\nu(k)\int {d^{4}p\over (2\pi)^{4}}
{\rm tr}\left[\sigma^\mu S_{ra}(p+k)\sigma^\nu S_{rr}(p)
+\sigma^\mu S_{rr}(p)\sigma^\nu S_{ar}(p-k)\right]\,.\label{Jres}
\ee

What we would like to do is to include the damping rate in the propagators, which is responsible for relaxation dynamics similar to the collision term in Boltzmann equation in relaxation time approximation.
One cautionary remark is that they don't represent the precisely same dynamics, and the order of the magnitude of the damping rate and that of the inverse of the relaxation time are in fact different in gauge theory~\cite{Blaizot:1999xk}:
the former is of order $e^2T \ln(1/e)$~\cite{Blaizot:1996az, Blaizot:1996hd, Lebedev:1990kt, Lebedev:1990un, Pisarski:1993rf} while the latter is $e^4T\ln(1/e)$.
In Eq.~(\ref{Jres}), we use the following propagators, in which the damping rate of the fermion ($\zeta/2$) is resummed~\cite{ValleBasagoiti:2002ir,Gagnon:2006hi}:
\bear 
S_{ra}(p)&=&\sum_{s=\pm}{i\over p^0-s|\vp|+i\zeta/2}{\cal P}_s(\vp)
\,,\label{newra}\\
S_{ar}(p)&=&\sum_{s=\pm}{i\over p^0-s|\vp|-i\zeta/2}{\cal P}_s(\vp)
\,\label{newar},\\
S_{rr}(p)&=&\left({1\over 2}-n_+(p^0)\right)\rho(p)\,,\label{newrr}
\eear
where the spectral density $\rho$ is 
\be
\rho(p)=\sum_{t=\pm}{\zeta\over (p^0-t|\vp|)^2+(\zeta/2)^2}{\cal P}_t(\vp)
\,.
\ee
Here the projection operators ${\cal P}_\pm(\vp)$ are defined as,
\be
{\cal P}_\pm(\vp)\equiv{1\over 2}\left({\bf 1}\pm {\vsigma\cdot\vp\over |\vp|}\right)
= \mp\frac{\bar\sigma\cdot p_\pm}{2|\vp|}\,,
\ee
where $\bar\sigma^\mu=({\bf 1},-\vsigma)$ and $p_\pm\equiv (\pm|\vp|,\vp)$.
The operators ${\cal P}_\pm(\vp)$ project onto particle and anti-particle states respectively with given momentum $\vp$. 
By using these expressions, Eq.~(\ref{Jres}) becomes
\begin{align}
\label{eq:Jres-2}
\begin{split}
j^\mu(k)&=e^2A_\nu(k)\int {d^{4}p\over (2\pi)^{4}}\sum_{s,t=\pm} \left({1\over 2}-n_+(p^0)\right) {\zeta\over (p^0-t|\vp|)^2+(\zeta/2)^2} \\
&~~~\times{\rm tr}\left[ {\sigma^\mu{\cal P}_s(\vp+\vk)
\sigma^\nu {\cal P}_t(\vp)\over p_+ +\omega-s|\vp+\vk|} 
+{\sigma^\mu {\cal P}_t(\vp)\sigma^\nu {\cal P}_s(\vp-\vk)\over p_- -\omega-s|\vp-\vk|}\right]
\,,
\end{split}
\end{align}
where we have introduced $p_\pm\equiv p^0\pm i\zeta/2$.
We note that the dominant contribution comes from the region $|\vp|\sim T, \mu$, so $\zeta\ll |\vp|$ is satisfied.
To compute the spinor trace, we use the following formula in the calculation,
\be
\label{eq:trace}
{\rm tr}\left[\sigma^\mu\bar\sigma^\alpha\sigma^\nu\bar\sigma^\beta\right]
= 2(g^{\mu\alpha}g^{\nu\beta}+g^{\mu\beta}g^{\nu\alpha}-g^{\mu\nu}g^{\alpha\beta})
+2i\epsilon^{\mu\alpha\nu\beta}.
\ee
The first term in the right-hand side gives the P-even contribution while the second term gives the P-odd one.
Though we are interested in the P-odd contribution, we will calculate both of them, to get a clear contrast on the absence of the pinch singularity in the P-odd one.

\subsection{P-even part}
First, let us obtain the P-even part.
We focus on the longitudinal component, to obtain the expression for $\sigma$.
By assuming that $\vk=k\hat{z}$, the longitudinal component of the current ($\mu=\nu=3$) reads
\begin{align}
\begin{split}
j^3(k)&= e^2A_3(k)\int {d^{4}p\over (2\pi)^{4}}\sum_{s,t=\pm} \frac{st}{2|\vp|^2}\left({1\over 2}-n_+(p^0)\right) {\zeta\over (p^0-t|\vp|)^2+(\zeta/2)^2} \\
&~~~\times\left[ {1\over 
p_+ +\omega-s|\vp|} 
+{1\over p_- -\omega-s|\vp|}\right](2(p_3)^2-\vp^2+st|\vp|^2) \\
&= -\frac{e^2A_3(k)}{\pi^2} \int^\infty_{-\infty} \frac{dp^0}{2\pi} \int^\infty_0 d|\vp| |\vp|^2  
\left({1\over 2}-n_+(p^0)\right) \\
&~~~\times{\zeta\over [|\vp|+p_+][|\vp|+p_-][|\vp|-p_-][|\vp|-p_+]} \\
&~~~\times\left[ 
{(p_++\omega)(|\vp|^2+p_+p_-)-2|\vp|^2p^0/3\over 
[|\vp|+p_++\omega][|\vp|-(p_++\omega)]} 
+{(p_--\omega)(|\vp|^2+p_+p_-)-2|\vp|^2p^0/3\over
 [|\vp|+p_--\omega][|\vp|-(p_--\omega)]}\right]
\,,
\end{split}
\end{align}
where we have taken $k=0$ (see Eq.~(\ref{eq:Kubo-sigma})).

Since we are interested in $\sigma$, we expand this expression in terms of $\omega$.
The zeroth order terms are
\begin{align}
\begin{split}
j^3(k)&= -\frac{e^2A_3(k)}{\pi^2} \int^\infty_{-\infty} \frac{dp^0}{2\pi} \int^\infty_0 d|\vp| |\vp|^2  
\left({1\over 2}-n_+(p^0)\right) \\
&~~~\times{\zeta\over [|\vp|+p_+][|\vp|+p_-][|\vp|-p_-][|\vp|-p_+]} \\
&~~~\times\left[ 
{p_+(|\vp|^2+p_+p_-)-2|\vp|^2p^0/3\over 
[|\vp|+p_+][|\vp|-p_+]} 
+{p_-(|\vp|^2+p_+p_-)-2|\vp|^2p^0/3\over
 [|\vp|+p_-][|\vp|-p_-]}\right]
\,.
\end{split}
\end{align}
This expression vanishes after we do the $|\vp|$ integration.
The first order terms,
\begin{align}
\begin{split}
j^3(k)&= -\omega\frac{e^2A_3(k)}{\pi^2} \int^\infty_{-\infty} \frac{dp^0}{2\pi} \int^\infty_0 d|\vp| |\vp|^2  
\left({1\over 2}-n_+(p^0)\right) \\
&~~~\times{\zeta\over [|\vp|+p_+][|\vp|+p_-][|\vp|-p_-][|\vp|-p_+]} 
\biggl[ \frac{(p^2_+-p^2_-)(|\vp|^2+p_+p_-)}{(|\vp|^2-p^2_+)(|\vp|^2-p^2_-)}\\
&~~~+2p_+\frac{p^2_+p_-+|\vp|^2(2p_+-p_-)/3}{(|\vp|^2-p^2_+)^2}
-2p_-\frac{p_+p^2_-+|\vp|^2(2p_--p_+)/3}{(|\vp|^2-p^2_-)^2}
\biggr]
\,,
\end{split}
\end{align}
can be calculated by using the residue theorem, which results in
\begin{align}
\begin{split}
j^3(k)&= \omega\frac{e^2A_3(k)}{3i\zeta\pi^2}
\int^\infty_{-\infty} dp^0 p^0 \left({1\over 2}-n_+(p^0)\right) \,.
\end{split}
\end{align}
By using 
$\int^\infty_{-\infty} dp^0 p^0 \left({1/ 2}-n_+(p^0)\right)
= \int^\infty_{0} dp^0 p^0 \left(1-n_+(p^0)-n_-(p^0)\right)$, we get
\begin{align}
\begin{split}
j^3(k)&= i\omega\frac{e^2A_3(k)}{3\zeta}\chi
\,,
\end{split}
\end{align}
where we have neglected the contribution from $T=\mu=0$ part, which corresponds to subtracting the vacuum contribution.
Comparing with Eq.~(\ref{eq:j-longitudinal}) at $k=0$, we have
\begin{align}
\begin{split}
\sigma&= {e^2\chi\over 3\zeta}
\,.
\end{split}
\end{align}
This result shows that we reproduce the expression of the conductivity obtained in the kinetic equation with the relaxation time approximation Eq.~(\ref{eq:sigma-kinetic}), by identifying $\zeta=\tau^{-1}$.
Note the $1/\zeta$ behavior of $\sigma$, which indicates the typical pinch singularity of the current-current correlation function at zero frequency-momentum limit.
In a theory where the chiral fermion interacts via gauge field with small coupling constant, the damping rate $\zeta/2$ is of order $e^2T\log(1/e)$~\cite{Blaizot:1996az, Blaizot:1996hd, Lebedev:1990kt, Lebedev:1990un, Pisarski:1993rf} and the $1/\zeta\sim 1/(e^2T\log(1/e))$ behavior in $\sigma$ due to the pinch singularity introduces non-analytic dependence of various transport coefficients on the coupling constant $e$, and the modified power counting necessitates the resummation of infinite ladder diagrams in the vertex function \cite{Jeon:1994if,Jeon:1995zm}.
However, we will see shortly that for our P-odd part of the correlation function responsible for the CME, the story will be different.

\subsection{P-odd part}
Next, we calculate the P-odd part.
For this part, we need to consider the case of $\mu=i, \nu=j$ only in (\ref{eq:Jres-2}).
By using the second term of Eq.~(\ref{eq:trace}), Eq.~(\ref{eq:Jres-2}) becomes
\begin{align}
\label{eq:Jres-odd}
\begin{split}
j^i(k)&=i\epsilon^{ijl}e^2A_j(k)\int {d^{4}p\over (2\pi)^{4}}\sum_{s,t=\pm} \frac{st}{2|\vp|}\left({1\over 2}-n_+(p^0)\right) {\zeta\over (p^0-t|\vp|)^2+(\zeta/2)^2} \\
&~~~\times\left[ \frac{1}{|\vp+\vk|}
{t|\vp|(p+k)_l-sp_l|\vp+\vk|\over p_+ +\omega-s|\vp+\vk|} 
+\frac{1}{|\vp-\vk|}
{s|\vp-\vk|p_l-t|\vp|(p-k)_l\over p_- -\omega-s|\vp-\vk|}\right] \\
&= i\epsilon^{ijl}e^2A_j(k)\int {d^{4}p\over (2\pi)^{4}} \left({1\over 2}-n_+(p^0)\right) 
{2\zeta\over  [|\vp|+p_+][|\vp|+p_-][|\vp|-p_-][|\vp|-p_+]} \\
&~~~\times\Bigl[ 
{(|\vp|^2+p_+p_-)(p+k)_l
-2p^0p_l (p_+ +\omega)\over (p_+ +\omega)^2-|\vp+\vk|^2} \\
&~~~+{2p^0p_l(p_- -\omega)
-(|\vp|^2+p_+p_-)(p-k)_l\over (p_- -\omega)^2-|\vp-\vk|^2}\Bigr]
\,,
\end{split}
\end{align}
where $l$ is different from $i$ and $j$. 
From the above expression, it is clear that the small $\omega, k\to 0$ limit is smooth due to the existence of the damping rate $\zeta$: 
the denominators in the integrand never vanish for real values of $\omega$, $\vk$, $p^0$ and $\vp$, so that the result can't have any singularity around $\omega, k\to 0$. 
This demonstrates that there is no non-commutativity issue between the two limits, $\omega\to 0$ and $k\to 0$.

Since we are interested in the momentum region $\omega, k\to 0$, we expand the expression in terms of $k$.
The zeroth order term reads
\begin{align}
\begin{split}
j^i(k)&= i\epsilon^{ijl}e^2A_j(k)\int {d^{4}p\over (2\pi)^{4}} \left({1\over 2}-n_+(p^0)\right) 
{2\zeta p_l\over [|\vp|+p_+][|\vp|+p_-][|\vp|-p_-][|\vp|-p_+]} \\
&~~~\times\Bigl[ 
{|\vp|^2+p_+p_-
-2p^0 p_+\over (p_+)^2-|\vp|^2} 
+{2p^0 p_-
-(|\vp|^2+p_+p_-)\over (p_-)^2-|\vp|^2}\Bigr]
\,.
\end{split}
\end{align}
By using $\zeta\ll p$, this expression seems to be of order $e^2A^\mu T\zeta $.
However, we see that it vanishes after angular integration.
The first order term is
\begin{align}
\begin{split} 
j^i(k)&= i\epsilon^{ijl}e^2A_j(k)k_l\int {d^{4}p\over (2\pi)^{4}} \left({1\over 2}-n_+(p^0)\right) 
{2\zeta\over [|\vp|+p_+][|\vp|+p_-][|\vp|-p_-][|\vp|-p_+]} \\
&~~~\times\Biggl[ 
{|\vp|^2+p_+p_-
\over (p_+)^2-|\vp|^2} 
+2|\vp|^2\cos^2\theta{|\vp|^2+p_+p_-
-2p^0 p_+\over [(p_+)^2-|\vp|^2]^2} \\
&~~~+{|\vp|^2+p_+p_-\over (p_-)^2-|\vp|^2}
+2|\vp|^2\cos^2\theta{|\vp|^2+p_+p_--2p^0p_-
\over [(p_-)^2-|\vp|^2]^2}\Biggr] \\
&=  -\frac{i\epsilon^{ijl}e^2}{2\pi^2}A_j(k)k_l 2\zeta
\int^\infty_{-\infty} \frac{dp^0}{2\pi}\left({1\over 2}-n_+(p^0)\right) 
\int^\infty_0 d|\vp||\vp|^2 \left[{|\vp|^2\over 3}+p_+p_-\right]  \\
&~~~\times\Biggl[ 
{1\over [|\vp|+p_+]^2[|\vp|+p_-][|\vp|-p_-][|\vp|-p_+]^2} \\
&~~~+{1\over [|\vp|+p_+][|\vp|+p_-]^2[|\vp|-p_-]^2[|\vp|-p_+]}\Biggr] 
\,,
\end{split}
\end{align}
where $\theta$ is an angle between $\vp$ and $\vk$.
By using the residue theorem, the $|\vp|$ integration can be explicitly performed to give
\begin{align}
\begin{split} 
j^i(k)&=  -\frac{i\epsilon^{ijl}e^2}{4\pi^2}A_j(k)k_l 
\int^\infty_{-\infty} dp^0\left({1\over 2}-n_+(p^0)\right) \\
&=  \frac{i\epsilon^{ijl}e^2\mu}{4\pi^2}A_j(k)k_l \,.
\end{split}
\end{align}
In going from the first line to the second, we used 
\bear
&&\int_{-\infty}^{\infty} dp^0 \left({1\over 2}-n_+(p^0)\right)
=-\int_{0}^{\infty} dp^0\,\left(n_+(p^0)-n_-(p^0)\right)=-\mu\,,
\eear
independent of temperature, which shows the contributions from
particles ($n_+(p^0)$) and anti-particles ($n_-(p^0)$) in a democratic way.

This expression is the one predicted from the anomaly relation, and agrees with the result obtained by the kinetic equation in $\omega$, $k\to 0$ limit.
We note that we can safely take both $\omega\to 0$ and $k\to 0$ limits, and both results agree.
We also emphasize that, this expression does not depend on $\zeta$.
This property reflects that the CME current is not renormalized by interaction.
It also implies that, we can take $\zeta\to 0$ limit, which means that there is no pinch singularity in the P-odd part, in contrast to the P-even part.
This indicates that the power counting for P-odd part of the correlation functions may be very different from (much simpler than) that of the P-even part, and one may not need to include infinite ladder summation.

\section{Summary and Concluding Remarks}
\label{sec:summary}
In this work, we argue that the non-commutativity between the two limits, $\omega\to 0$ and $k\to 0$, disappears in any interacting theory whose low-energy behavior is described by hydrodynamics.
We confirm our argument at weak coupling regime in the frameworks of the Boltzmann equation in the relaxation time approximation and the diagrammatic method with the resummation of the damping rate. We reproduce the correct magnitude of the CME current in both formalisms.

We comment that the situation may be different for chiral vortical effect whose Kubo formula involves both the current and energy-momentum, due to
a possible coupling to the diffusive pole of the energy-momentum sector, as recently observed in Ref.\cite{Landsteiner:2013aba} \footnote{We thank Yi Yin for pointing this out to us.}. A related subtlety in our discussion is a possible appearance of new diffusive pole in our P-odd current-current response function
if we included the coupling to the energy-momentum sector. Independently to this, however, it is our robust conclusion that the non-commutativity of the two limits observed in previous literature is an artifact of the free theory limit, which disappears with finite interactions.

We observe the absence of pinch singularity in the P-odd part of correlation function at zero frequency-momentum limit, which is in sharp contrast to the P-even part. Recalling that the pinch singularity is a signal for the need of resummation of infinite ladder diagrams to get the leading order result in weak coupling, this may indicate a different power counting scheme for the P-odd part. 
We hope to address this issue in detail in the near future.

\vskip 1cm \centerline{\large \bf Acknowledgment} \vskip 0.5cm

We thank Dima Kharzeev, Shu Lin, Dam T. Son, Misha Stephanov, and Yi Yin for helpful discussions and comments.
D. S. is supported by JSPS Strategic Young Researcher Overseas Visits Program for Accelerating Brain Circulation (No. R2411).

\vfil

\end{document}